\documentclass[notitlepage,prd,showpacs,superscriptaddress,nofootinbib,floatfix,showkeys,10pt]{revtex4-1}
\usepackage{graphicx}
\usepackage{amsmath}
\usepackage{bm}
\usepackage{yhmath}
\usepackage{mathtools}
\usepackage{wasysym}
\usepackage{subfigure}
\usepackage{xcolor}
\definecolor{CherryRed}{rgb}{.65,0,.2}
\definecolor{RubyRed}{rgb}{.88,0.07,.3}
\definecolor{CralRed}{rgb}{1,0.25,.25}
\definecolor{CobaltBlue}{rgb}{0,0.28,.67}
\definecolor{RoyalBlue}{rgb}{0.25,0.41,.88}
\definecolor{EmeraldGreen}{rgb}{0.31,0.78,.47}
\usepackage{cases}
\usepackage[colorlinks=true, linkcolor=RubyRed,citecolor=CralRed,urlcolor=RubyRed,linkcolor=RubyRed]{hyperref}
\usepackage{subfigure}
\usepackage{times}
\usepackage{booktabs,bm}
\usepackage{slashed}
\usepackage{amsfonts,amssymb,stmaryrd,latexsym,amsmath}
\usepackage{textcomp}
\usepackage{multirow}
\usepackage{cancel}
\usepackage{array}
\usepackage{hyperref}
\usepackage{pgf}
\usepackage{orcidlink}

\def\SU3{{\text{SU(3)}_{\rm F}}}

\def \pcs4338{{P_{\psi s}^\Lambda(4338)^0}}

\begin{document}
	
	\title{\textcolor{CobaltBlue}{Investigation of  triply heavy spin-3/2 baryons in their ground and excited states}}

	\author{Z.~Rajabi Najjar\orcidlink{0009-0002-2690-334X}}
	\email{rajabinajar8361@ut.ac.ir }
	\affiliation{Department of Physics, University of Tehran, North Karegar Avenue, Tehran 14395-547,  Iran }

	\author{K.~ Azizi\orcidlink{0000-0003-3741-2167}}
	\email{kazem.azizi@ut.ac.ir}
	\thanks{Corresponding author}
	\affiliation{Department of Physics, University of Tehran, North Karegar Avenue, Tehran 14395-547,  Iran }
	\affiliation{Department of Physics,  Dogus University, Dudullu-\"{U}mraniye, 34775 Istanbul, T\"urkiye }

	\date{\today}

	\begin{abstract}
       We calculate the masses and residues  of triply heavy baryons with spin-3/2, including  $\Omega^*_{ccc}$, $\Omega^*_{ccb}$, $\Omega^*_{bbc}$ and  $\Omega^*_{bbb}$, using the QCD sum rules method.  Our calculations primarily focus on obtaining the masses of the first three resonances, that is, the ground state (1S), the first orbital excited state (1P), and the first radial excited state (2S), for  the mentioned baryons. We additionally determine the residues of these baryons, which serve as key parameters for studying their possible decay channels and interactions with other particles. To achieve higher accuracy compared to  previous studies, we consider nonperturbative operators up to  eight mass dimensions. We present  our calculated outcomes in two distinct energy schemes, referred to as pole and $\mathrm{\overline{MS}}$. Given the absence of experimental data for these states, we compare our results with previous theoretical calculations that are reported in relevant studies employing various approaches. These results may provide valuable insights for experimental groups searching for the  triply heavy baryons.
      	\end{abstract}

	\maketitle
	\renewcommand{\thefootnote}{\#\arabic{footnote}}
	\setcounter{footnote}{0}


\section {Introduction}\label{sec:one}

One of the key topics in hadronic physics is the examination of the features of heavy baryons as predicted by the successful  quark model.  Among the hadronic states containing singly, doubly and triply heavy charm or bottom quarks, singly heavy baryons have been the focus of significant investigation in the literature, both theoretically and experimentally, as outlined in the summary tables issued by the Particle Data Group  (PDG)  \cite{ParticleDataGroup:2022pth}.  Due to the discovery and subsequent confirmation of   $\Xi_{cc}$ by the SELEX and LHCb collaborations \cite{SELEX:2002wqn,SELEX:2004lln, LHCb:2017iph,LHCb:2018pcs}, the previous decade has been highly productive in the investigation of doubly heavy baryon.  These observations have sparked renewed theoretical and experimental investigations into the properties of heavy baryons. However, no experimental evidence for  triply heavy baryons has been observed so far. Baranov \textit{et. al.}  predicted that the $\Omega^*_{ccc}$ might be undetectable  in $e^{+}e^{-}$ collisions, and the likelihood of observing the $\Omega^*_{bbb}$   is even more unlikely  \cite{Baranov:2004er}. First projections of the production cross-section for triply heavy baryons  at the LHC were evaluated in Refs. \cite{Saleev:1999ti,GomshiNobary:2003sf,GomshiNobary:2004mq,GomshiNobary:2005ur}. In 2011, it was projected that around $10^{4}-10^{5}$ events of some triply heavy baryons, namely $\Omega_{ccc}^*$, $\Omega_{ccb}^*$ and $\Omega_{ccb}$, could be accumulated with an integrated luminosity of 10 $fb^{-1}$  at the LHC \cite{Chen:2011mb}. The production rates of hadrons containing charm quark in high-energy heavy-ion collisions are examined in Refs. \cite{He:2014tga,Zhao:2017gpq,ExHIC:2017smd,Cho:2019syk}  presents a quantitative analysis of $\Omega^*_{ccc}$ production in heavy-ion collisions using the coalescence model, including predictions for its yield and transverse momentum distribution at RHIC and LHC energies.
In Ref. \cite{Li:2025nca}, the production of the $\Omega_{ccc}^*$ baryon in relativistic heavy-ion collisions is studied, with a focus on thermal production and coalescence mechanisms.
Based on these studies, there is a high probability of observing triply charmed baryons at the LHC in near future.

 Theoretical studies of triply heavy baryons could open a new avenue for deepening our understanding of hadron structure, the dynamics of the  c and b quarks at the hadronic energy  scale, and the symmetry of heavy quarks. They may also be utilized to analyze the effects of perturbative and nonperturbative terms. As a result, these investigations can offer invaluable insights and crucial data that will play a key role in guiding future experimental discoveries. In light of this, the broad range of theoretical methods  has been applied to study  the features of  baryons with three heavy quarks, which can exist in spin states of either 1/2 or 3/2, including  Bag model \cite{Hasenfratz:1980ka,Bernotas:2008bu}, lattice QCD \cite{Meinel:2010pw,Meinel:2012qz,Briceno:2012wt,Padmanath:2013zfa,Brown:2014ena,Mathur:2018epb,Can:2015exa},  non-relativistic quark model \cite{Roberts:2007ni,Patel:2008mv,Vijande:2015faa,Shah:2017jkr,Shah:2018div,Shah:2018bnr,Liu:2019vtx,Ortiz-Pacheco:2023kjn}, relativistic quark model  \cite{Migura:2006ep,Martynenko:2007je,Yang:2019lsg,Faustov:2021qqf},  Faddeev equation \cite{Radin:2014yna,Gutierrez-Guerrero:2019uwa,Qin:2019hgk,Yin:2019bxe,Gutierrez-Guerrero:2024him,Sanchis-Alepuz:2011xjl}, Hyperspherical Harmonics method \cite{Zhao:2023qww},  QCD sum rules \cite{Zhang:2009re,Wang:2011ae,Aliev:2012tt,Aliev:2014lxa,Wang:2020avt,Najjar:2024deh},   Regge trajectories  \cite{Wei:2015gsa,Wei:2016jyk,Oudichhya:2021yln,Oudichhya:2021kop,Oudichhya:2023pkg,Xie:2024lfo}, various potential models  \cite{Ghalenovi:2011zz,Silvestre-Brac:1996myf,Jia:2006gw,Brambilla:2009cd,Llanes-Estrada:2011gwu,Flynn:2011gf,Thakkar:2016sog,Serafin:2018aih,Shah:2023zph}, etc. Among the various methods discussed, the QCD sum rule approach is regarded as a highly efficient nonperturbative formalism for predicting  the properties of heavy hadrons \cite{Aliev:2009jt,Aliev:2010uy,Aliev:2012ru,Agaev:2016mjb,Azizi:2016dhy}. Given that the predictions from this method have been validated by numerous accelerator experiments globally, substantial research in the literature has been dedicated to investigating the properties of multi-heavy-quark hadrons within the framework of QCD sum rules.

To date, theoretical investigations have primarily focused on the ground-state triply heavy baryons, particularly their masses, while comparatively less attention has been given to their excited states. Therefore, to gain a comprehensive understanding of the excitation spectrum, further in-depth theoretical investigations are necessary to thoroughly characterize the properties and behaviors of the excited states of triply heavy baryons, which may exist in spin-1/2 or spin-3/2 configurations. Motivated by this need, in Ref. \cite{ Najjar:2024deh}, we investigated the spectral properties, including the masses and residues, of the ground state, along with the first orbitally and radially excited states of the triply heavy spin-1/2 baryons using the QCD sum rule approach. Building upon the promising outcomes of previous studies, we aim to calculate the mass and residue parameters of triply heavy baryons with spin-3/2 in the 1S, 1P,  and 2S states, utilizing the QCD sum rule approach in both the pole and $\mathrm{\overline{MS}}$ schemes for heavy quarks. In prior studies utilizing the sum rule method \cite{Aliev:2014lxa}, the evaluation of physical quantities, including mass and residue, was limited to four mass-dimensional operators for both the ground state (1S) and the first orbitally excited state (1P) of baryons containing three heavy quarks with spin-3/2.  In order to increase accuracy through the use of  contributions from higher-dimensional nonperturbative operators, we extend our calculations to include nonperturbative terms up to dimension eight for the ground (1S) state, the first orbital excited (1P) state  and the first radial excited (2S) state. These findings can be utilized in subsequent experiments at research facilities, including those at the LHC, to explore various species of heavy baryons.

The structure of this paper is outlined below: Section \ref{sec:two}  presents a detailed formulation of the QCD sum rule technique to extract the mass spectra and residue values for the triply heavy  baryons with spin-3/2 across their first three resonances (FTRs), while Section  \ref{sec:three} presents the outcomes of the numerical evaluation of the sum rules and compares these results with those reported in the current literature. Section \ref{sec:four} provides a brief synopsis and concluding remarks on the investigation. Detailed expressions derived from the calculations are provided in the Appendix.

\section { THEORETICAL FRAMEWORK }\label{sec:two}

The QCD sum rule method enables the determination of spectroscopic parameters, including mass and residue \cite{Shifman:1978bx,Shifman:1978by,Ioffe:1981kw}. To establish the related sum rules for triply heavy baryons with spin-3/2, an appropriate correlation function should be utilized. The correlation function presented below serves as the starting point for the evaluation in this study
\begin{equation}
		\Pi_{\mu\nu}(q)=i\int d^{4}xe^{iqx}\langle 0|\mathcal{T}\{\eta_\mu (x)\bar{\eta}_\nu (0)\}|0\rangle, 
		 \label{eq:CF1}
	\end{equation}
	where $\mathcal{T}$ denotes the time ordering operator, $q$ is the four-momentum for the triply heavy  baryons with spin-3/2, and $\eta_\mu$ is used to represent  interpolating current of these baryons, which can  generally be expressed as:	
	\begin{equation}\label{cur1}
		\eta_\mu = {1\over \sqrt{3}} \epsilon^{abc} \Big\{2
(Q^{aT} C \gamma_\mu Q^{\prime b}) Q^c +
(Q^{aT} C \gamma_\mu Q^b) Q^{\prime c} \Big\}.
	\end{equation}
	In the above interpolating current expression,  $Q$ and $Q' $  denote the heavy quarks ($b$ or $c$ ), $C$ is charge conjugation operator. The indices $a$, $b$ and $c$  denote the color indices of the heavy quark fields. The presence of the fully antisymmetric tensor $\epsilon^{abc}$ ensures that the baryonic current is a color singlet and guarantees the antisymmetry of the color wave function, as required by the Pauli principle. Table~\ref{tab1}  presents the quark contents for all triply heavy  baryons with spin-3/2, as predicted by the quark model.	
	\begin{table}[htb]
		\begin{center}
			\begin{tabular}{|c|c|c|c|c|}\hline\hline
				Baryon        & $\Omega^*_{ccc}$ &  $\Omega^*_{ccb}$ & $\Omega^*_{bbc}$ & $\Omega^*_{bbb}$  \\ \hline
      $Q$ &  $c$           &$c$  &$b$   &$b$\\ \hline
      $Q' $ &  $c$           &$b$  &$c$ & $b$\\ \hline
			\end{tabular}
		\end{center}
		\caption{The members of the spin-3/2 baryon family, consisting of three heavy quarks.}
		\label{tab1}
	\end{table}
	The primary objective in the QCD sum rule method is to calculate the two-point correlation function, Eq (\ref{eq:CF1}), within two distinct regimes known as the hadronic and QCD sides. At large distances, within the time-like region,  the calculated correlation function depends on the hadronic parameters:  mass and residue. This aspect is referred to as the hadronic or physical description of  the correlation function. In contrast, the identical two-point correlation function is analyzed within the Operator Product Expansion (OPE) framework through various dimensions of the gluon condensates at $q^2\ll\ 0 $, which corresponds to the   space-like  region.  This computational scheme  includes short-distance effects. This side is commonly referred to as the OPE or QCD representation  of the correlation function. By constructing two representations of a correlation function, we relate them using a dispersion relation and the quark-hadron duality assumption to formulate the QCD sum rules for the  desired physical quantities.  Unwanted contributions from higher states and the continuum are suppressed on both sides by applying Borel transformation and continuum subtraction operations. The results obtained from both sides include various Lorentz structures.  To extract the desired physical parameters, it is necessary to match the coefficients of the corresponding Lorentz structures.
	 \subsection{Hadronic side} 
	 	On the hadronic side, considering the interpolating current as the operator for creation and annihilation from vacuum, the studied states are generated or annihilated. To derive this side of the correlation function, the QCD sum rule framework starts  systematically by inserting the two-point correlation function with complete sets of baryonic states in which the quark content has the same quantum numbers as the interpolating current:	
	\begin{equation}
    	1\,=\,|\,0\rangle\,\langle0\,|+\sum_{h}\,\int\,\frac{\mathrm{d}^{4}p_{h}}{(2\pi)^{4}}\,2\pi\,\delta(p_{h}^{2}-m_{h}^{2})|\,h(p_{h})\,\rangle\,\langle\,h(p_{h})\,|\,\,+\,\text{higher Fock states},
    \end{equation}
	where $|\,h(p_{h})\rangle$ represents the considered hadronic state with four-momentum  $p_{h}$. Subsequently, by doing the Fourier transformation through integrating over four-x and then separating the ground state  as well as the corresponding excited states, we get the explicit form of the hadronic correlation function:
	\begin{eqnarray}
		\Pi^{\mathrm{Had}}_{\mu \nu}(q)&=&\frac{\langle0|\eta_{\mu}| \Omega^*_{QQQ^{'}}(q,s)\rangle\langle \Omega^*_{QQQ^{'}}(q,s)|\bar{\eta}_{\nu}|0\rangle}{m^2-q^2}
		+\frac{\langle0|\eta_{\mu}|\tilde{\Omega}^*_{QQQ^{'}}(q,s)\rangle\langle\tilde{\Omega}^*_{QQQ{'}}(q,s)|\bar{\eta}_{\nu}|0\rangle}{\tilde{m}^2-q^2}+\nonumber\\
		&+&\frac{\langle0|\eta_{\mu}| \Omega^{*'}_{QQQ{'}}(q,s)\rangle\langle \Omega^{*'}_{QQQ{'}}(q,s)|\bar{\eta}_{\nu}|0\rangle}{m'{}^2-q^2}+\cdots,
		\label{Eq:cor:Phys}
	\end{eqnarray}	
	where the terms $| \Omega^*_{QQQ^{'}}(q,s)\rangle$, $|\tilde{\Omega}^*_{QQQ^{'}}(q,s)\rangle$ and $|\Omega^{*'}_{QQQ^{'}}(q,s)\rangle$ represent the contributions of the one-particle spin-3/2 states of the 1S, 1P and 2S states, respectively. Here,  $m$, $\tilde{m}$ and $m'$  denote their respective masses, and dots are used to indicate the contributions of the higher states and continuum.  To proceed in the calculation of the matrix elements in Eq.~(\ref{Eq:cor:Phys}), we introduce:
\begin{eqnarray}
		\langle 0|\eta_{\mu}|\Omega^*_{QQQ^{'}}(q,s)\rangle&=&\lambda_{(\frac{3}{2})} u_{\mu}(q,s),\nonumber\\
		\langle 0|\eta_{\mu}|\tilde{\Omega}^*_{QQQ^{'}}(q,s)\rangle&=&\tilde{\lambda}_{(\frac{3}{2})} \gamma_5 u_{\mu}(q,s),\nonumber\\
		\langle 0|\eta_{\mu}|\Omega^{*'}_{QQQ^{'}}(q,s)\rangle&=&\lambda'_{(\frac{3}{2})} u_{\mu}(q,s),
				\label{Eq:Matrixelm}
	\end{eqnarray}  
	where    $\lambda_{(\frac{3}{2})}$, $\tilde{\lambda}_{(\frac{3}{2})}$ and $\lambda'_{(\frac{3}{2})}$  denote the corresponding residues and $u_{\mu}(q,s)$  represents the  Rarita–Schwinger spinor with spin $s$. In order to continue, it is necessary to sum over the Rarita-Schwinger spinors for spin-3/2 baryons, which takes the following form:
		\begin{eqnarray}
		\sum u_\mu(q,s) \bar{u}_\nu (q,s) &=& -(\!\not\!{q} + m) \Bigg( g_{\mu\nu} - {1
\over 3} \gamma_\mu \gamma_\nu - {2 q_\mu q_\nu \over 3 m_B^2} + {q_\mu
\gamma_\nu - q_\nu \gamma_\mu \over 3 m_B} \Bigg).
				\label{Eq:Summation}
	\end{eqnarray} 

	 Before proceeding to calculate the correlation function on the hadronic side, it is essential to note that the interpolating current of the triply heavy spin-3/2  baryons couples not only with spin-3/2 baryons but also with  triply heavy spin-1/2  baryons.  By imposing the condition $\gamma^{\mu} J_{\mu}=0$, the unwanted contributions from the spin-1/2 states can be expressed as: 
	\begin{eqnarray}
		\langle 0|\eta_{\mu}|\Omega_{QQQ^{'}}(q,s)\rangle&=&\lambda_{(\frac{1}{2})} \Bigg( {4 q_\mu \over m_{(\frac{1}{2})^+}} + \gamma_\mu \Bigg) \gamma_5 u(q,s),\nonumber\\
		\langle 0|\eta_{\mu}|\tilde{\Omega}_{QQQ^{'}}(q,s)\rangle&=&\tilde{\lambda}_{(\frac{1}{2})} \Bigg( {- 4 q_\mu \over \tilde{m}_{(\frac{1}{2})^-}} +
\gamma_\mu \Bigg)  u(q,s),\nonumber\\
		\langle 0|\eta_{\mu}|\Omega_{QQQ^{'}}'(q,s)\rangle&=&\lambda'_{(\frac{1}{2})}  \Bigg( {4 q_\mu \over m'_{(\frac{1}{2})^+}} + \gamma_\mu \Bigg) \gamma_5 u(q,s),
				\label{Eq:Matrixelm2}
	\end{eqnarray} 	
	where    $\lambda_{(\frac{1}{2})}$, $\tilde{\lambda}_{(\frac{1}{2})}$ and $\lambda'_{(\frac{1}{2})}$  are utilized to represent the various spin-1/2 residues: the 1S, 1P and 2S states, respectively. $u(q,s)$  is the  Dirac spinor. Based on the matrix elements in Eqs.~(\ref{Eq:Matrixelm2}), the Lorentz structures, including $\gamma_{\mu}$ and $q_{\mu}$, also receive contributions from spin-1/2 states. To eliminate these undesirable contributions mixed with spin-3/2 states, it is necessary to select structures that exclusively receive contributions from spin-3/2 states. 	By applying the aforementioned relations, the outcome on the hadronic side is obtained as follows:
	\begin{eqnarray}
	\label{PhysSide1}
\Pi_{\mu\nu}^{\mathrm{Had}}(q)&=&\frac{\lambda_{(\frac{3}{2})}^{2}}{m_{(\frac{3}{2})}^{2}-q^{2}}(\!\not\!{q} + m_{(\frac{3}{2})})\Big[g_{\mu\nu} -\frac{1}{3} \gamma_{\mu} \gamma_{\nu} - \frac{2q_{\mu}q_{\nu}}{3m_{(\frac{3}{2})}^{2}} +\frac{q_{\mu}\gamma_{\nu}-q_{\nu}\gamma_{\mu}}{3m_{(\frac{3}{2})}} \Big]\nonumber \\
&+&\frac{\tilde{\lambda}_{(\frac{3}{2})}^{2}}{\tilde{m}_{(\frac{3}{2})}^{2}-q^{2}}(\!\not\!{q} -\tilde{m}_{(\frac{3}{2})})\Big[g_{\mu\nu} -\frac{1}{3} \gamma_{\mu} \gamma_{\nu} - \frac{2q_{\mu}q_{\nu}}{3\tilde{m}_{(\frac{3}{2})}^{2}} -\frac{q_{\mu}\gamma_{\nu}-q_{\nu}\gamma_{\mu}}{3\tilde{m}_{(\frac{3}{2})}} \Big]\nonumber\\
&+&\frac{\lambda'{}^{2}}{m'{}^{2}_{(\frac{3}{2})}-q^{2}}(\!\not\!{q} + m'_{(\frac{3}{2})})\Big[g_{\mu\nu} -\frac{1}{3} \gamma_{\mu} \gamma_{\nu} - \frac{2q_{\mu}q_{\nu}}{3m'{}^{2}_{(\frac{3}{2})}} +\frac{q_{\mu}\gamma_{\nu}-q_{\nu}\gamma_{\mu}}{3m'_{(\frac{3}{2})}} \Big]+\cdots.
\end{eqnarray}	
It can be seen that this expression contains several Lorentz structures. However, only the $g_{\mu\nu}$ and $\!\not\!{q} g_{\mu\nu}$  structures are free from contamination by spin-1/2 and arise solely from baryons with spin-3/2.  Therefore, the final result of the hadronic side is as follows:
\begin{eqnarray}
\Pi _{\mu \nu}^{\mathrm{Had}}(q)&=&\frac{\lambda{}^2}{m^{2}-q^{2}}  (\!\not\!{q} + m)g_{\mu\nu}+
\frac{\tilde{\lambda}{}^2}{\tilde{m}{}^{2}-q^{2}}  (\!\not\!{q} -\tilde{m})g_{\mu\nu} +\frac{\lambda'{}^2}{m'{}^{2}-q^{2}} (\!\not\!{q} + m') g_{\mu\nu}+\cdots,
\label{eq:CorFun1}
\end{eqnarray}
where we omitted the sub-index   $\frac{3}{2}  $ for simplicity. 

The hadronic correlation function, after applying the Borel transformation with respect to $-q^2$ to suppress the contributions of higher and continuum states, is presented in its final form as follows:
\begin{eqnarray}
\mathcal{\widehat B}\Pi _{\mu \nu}^{\mathrm{Had}}(q)&=&\lambda^2 e^{-\frac{m^{2}}{M^{2}}} (\!\not\!{q}+ m) g_{\mu\nu}+
\tilde{\lambda}{}^2 e^{-\frac{\tilde{m}^{2}}{M^{2}}}  (\!\not\!{q} -\tilde{m})g_{\mu\nu}+\lambda'{}^2 e^{-\frac{m'^{2}}{M^{2}}} (\!\not\!{q} + m')g_{\mu\nu} +\cdots.
\label{eq:CorFunBorel}
\end{eqnarray}
 \subsection{QCD side} 
In this sector, the two-point correlator is evaluated in the deep space-like Euclidean domain. To obtain this quantity, the interpolating current of triply heavy baryons with spin-3/2, as presented in Eq.~(\ref{cur1}),  should be substituted into Eq.~(\ref{eq:CF1}), and all possible contractions of the heavy quark-antiquark fields are applied  using the Wick theorem.  As a result, a clear expression can be derived as a function of the heavy quark propagators:
\begin{eqnarray}
\label{QCDSide1}
\Pi_{\mu\nu} (q) &=& \frac13
\epsilon^{abc} \epsilon^{a'b'c'} \int d^4x
e^{iqx}  \Big\{ 4S_Q^{c b'} \gamma_\nu
S_{Q'}^{ ' b a'} \gamma_\mu S_Q^{a c'}
+2 S_Q^{c a'} \gamma_\nu S_{Q}^{' a b'}
\gamma_\mu S_{Q'}^{b c'} - 2S_{Q}^{c b'}
\gamma_\nu S_{Q}^{' a a'} \gamma_\mu
S_{Q'}^{b c'} \nonumber \\
&+& 2 S_{Q'}^{c a'} \gamma_\nu S_{Q}^{' a b'}
\gamma_\mu S_{Q}^{b c'} - 2 S_{Q'}^{c a'} \gamma_\nu
S_{Q}^{' b b'} \gamma_\mu S_Q^{a c'} -
S_{Q'}^{c c'}  \mbox{Tr}\Big[S_{Q}^{b a'} \gamma_\nu
S_{Q}^{' a b'} \gamma_\mu \Big] 
+
S_{{Q'}}^{c c'}  \mbox{Tr}\Big[S_{Q}^{b b'} \gamma_\nu
S_{Q}^{' a a'} \gamma_\mu \Big] \nonumber \\
&-&
4 S_{Q}^{c c'}  \mbox{Tr}\Big[S_{Q'}^{b a'} \gamma_\nu
S_{Q}^{' a b'} \gamma_\mu \Big] \Big\}.
\end{eqnarray}
Here,  $S_Q$  denotes the heavy quark propagator and   $S'_Q$  is defined as  $ C S_Q^T C$. The explicit formula for the heavy quark propagator in coordinate space is expressed as follows  \cite{Agaev:2020zad}:
\begin{eqnarray}\label{eqQCD2}
		&&S_{Q}^{ab}(x)=i\int \frac{d^{4}k}{(2\pi )^{4}}e^{-ikx}\Bigg \{\frac{\delta
			_{ab}\left( {\slashed k}+m_{Q}\right) }{k^{2}-m_{Q}^{2}}-\frac{%
			g_{s}G_{ab}^{\alpha \beta }}{4}\frac{\sigma _{\alpha \beta }\left( {\slashed %
				k}+m_{Q}\right) +\left( {\slashed k}+m_{Q}\right) \sigma _{\alpha \beta }}{%
			(k^{2}-m_{Q}^{2})^{2}}  \notag  \\
		&&+\frac{g_{s}^{2}G^{2}}{12}\delta _{ab}m_{Q}\frac{k^{2}+m_{Q}{\slashed k}}{%
			(k^{2}-m_{Q}^{2})^{4}}+\frac{g_{s}^{3}G^{3}}{48}\delta _{ab}\frac{\left( {%
				\slashed k}+m_{Q}\right) }{(k^{2}-m_{Q}^{2})^{6}}\left[ {\slashed k}\left(
		k^{2}-3m_{Q}^{2}\right) +2m_{Q}\left( 2k^{2}-m_{Q}^{2}\right) \right] \left(
		{\slashed k}+m_{Q}\right) +\cdots \Bigg \}.  \notag \\
		&&
	\end{eqnarray} 
Here $k$ represents  the four-momentum of the heavy quark, while $G^{\alpha \beta}$  denotes the gluon field strength tensor, expressed as  $G_{ab}^{\alpha \beta }=G_{A}^{\alpha \beta
}t_{ab}^{A}$ with $t^A=\lambda^A/2$.  Additionally, the notations $G^{2}=G_{A}^{\alpha \beta} G_{\alpha \beta }^{A}$  and  $ G^{3}=\,\,f^{ABC}G^{\alpha \beta }_{A}G^{ \beta
			\delta }_{B}G^{\delta \alpha  }_{C}$  are introduced.  The indices $\alpha$, $ \beta$ and $\delta$ correspond to Lorentz indices, while $\lambda^A$ represent the Gell-Mann matrices.  Furthermore, $f ^{ABC}$ denotes the structure constants of the color $SU_c(3)$ group, where A, B and C  range from 1 to 8. The correlation function on the QCD side can be expressed as perturbative and nonperturbative contributions with different mass dimensions. This can be done by inserting the  heavy quark propagator mentioned above, Eq.~(\ref{eqQCD2}),  into Eq.~(\ref{QCDSide1}).  In the calculations, we use nonperturbative operators up to the eight-mass dimension  that include glounic terms such as the two-gluon condensate, $\langle G^2 \rangle$; the three-gluon condensate, $\langle G^3 \rangle$, and the double two-gluon condensate, $\langle G^2 \rangle^2$, all of which are associated with the QCD vacuum and represent four-mass dimension, six-mass dimension and eight-mass dimension, respectively. The extraction of each OPE term, including both perturbative and nonperturbative contributions up to dimension eight, follows the comprehensive methodology presented in our previous work, Ref  \cite{Najjar:2024deh}, where all relevant diagrams were explicitly calculated and discussed in detail. These results are formulated based on Fourier integrals that can be calculated in a mathematical framework using various techniques (for more details  see Refs. \cite{Najjar:2024deh,Najjar:2024ngm}). As pointed out previously, there are two distinct Lorentz structures on  the QCD and hadronic sides , which are labeled by  $g_{\mu\nu}$ and $\not\!q g_{\mu\nu}$. Finally, the result on the QCD side,  using the  Borel transformations as well
as continuum subtraction, is expressed for two Lorentz structures as follows:
\begin{eqnarray}
		\Pi^{\mathrm{QCD}}_{\not\!q g_{\mu\nu}}(s_0,M^2)=\int_{(2m_Q+m_{Q'})^2}^{s_0}ds\,e^{-\frac{s}{M^2}}\rho_{\not\!q g_{\mu\nu}}(s)+\Gamma_{\not\!q g_{\mu\nu}}(M^2),
		\label{Eq:finalCor:QCD1}
	\end{eqnarray}
	and
	\begin{eqnarray}
		\Pi^{\mathrm{QCD}}_{ g_{\mu\nu}}(s_0,M^2)=\int_{(2m_Q+m_{Q'})^2}^{s_0}ds\,e^{-\frac{s}{M^2}}\rho_{ g_{\mu\nu}}(s)+\Gamma_{ g_{\mu\nu}}(M^2).
		\label{Eq:finalCor:QCD}
	\end{eqnarray}
	In this context, $s_0$ signifies the continuum threshold, while  $\rho _{\not\!q g_{\mu\nu}(g_{\mu\nu})}(s)=\frac{1}{\pi}\mathrm{Im}[\Pi_{\not\!q g_{\mu\nu}(g_{\mu\nu})}^{\mathrm{QCD}}]$ represents the spectral density, which  incorporates perturbative contributions. $\Gamma_{\not\!q g_{\mu\nu}(g_{\mu\nu})}(M^2) $ indicates a function with no imaginary parts that contains the nonperturbative operators: four, six and eight mass dimensions.  The detailed mathematical formulations corresponding to $\rho _{~\not\!q g_{\mu\nu}}(s)$ and certain components of $ \Gamma_{\not\!q g_{\mu\nu}}(M^2) $ are presented in the Appendix.   By equating the coefficients of the relevant structures on both the QCD and physical sides, we can derive sum rules for the mass and residue, as follows:

\begin{eqnarray}
	\lambda^2 e^{-\frac{m^2}{M^2}}+\tilde{\lambda}^2e^{-\frac{\tilde{m}^2}{M^2}}+\lambda'^2e^{-\frac{m'{}^2}{M^2}}=\Pi^{QCD}_{~\!\not\!{q} ~g_{\mu\nu}}(s_0,M^2),
	\label{Eq:cor:match1}
\end{eqnarray}
and
\begin{eqnarray}
	\lambda^2 m e^{-\frac{m^2}{M^2}}-\tilde{\lambda}^2\tilde{m}e^{-\frac{\tilde{m}^2}{M^2}}+\lambda'^2m'e^{-\frac{m'{}^2}{M^2}}=\Pi^{QCD}_{g_{\mu\nu}}(s_0,M^2).
	\label{Eq:cor:match2}
\end{eqnarray}
In the next step of this section, we will demonstrate how to extract the QCD sum rules for the masses and residues of the 1S, 1P, and 2S states for the structure $ \not\!q g_{\mu\nu}$.  To fulfill this aim, a comprehensive three-step method will be employed.  First, the mass and residue for the 1S state will be determined independently. Accordingly, we follow the 1S state $+$ continuum scheme, in which the terms other than the first on the left-hand side of Eq.~ (\ref{Eq:cor:match1}) are considered as continuum components. By adjusting the continuum threshold, $ s_0 $,  we can ensure that only the contribution from the 1S state appears in our calculations.  To obtain the mass and residue of 1S state, we need two equations, which would be found by using the derivative of  Eq.~ (\ref{Eq:cor:match1}) with respect to $ -\frac1{M^2} $. After some manipulations and by focusing solely on the first term on the left- hand side of Eq.~ (\ref{Eq:cor:match1}), the sum rules for the mass and residue of the 1S state can be extracted as follows: 
\begin{eqnarray}
		m^2=\frac{\frac{d}{d(-\frac{1}{M^2})}\Pi^{\mathrm{QCD}}_{~\not\! q~ g_{\mu\nu}}(s_0,M^2)}{\Pi^{\mathrm{QCD}}_{~\not\!q ~g_{\mu\nu}}(s_0,M^2)},
		\label{Eq:mass:Groundstates1}
	\end{eqnarray}
and
	\begin{eqnarray}
		\lambda^2=e^{\frac{m^2}{M^2}}\Pi^{\mathrm{QCD}}_{~\not\!q ~g_{\mu\nu}}(s_0,M^2).
		\label{Eq:residumass:Groundstates1}
	\end{eqnarray}

Similarly to the first step, the parameters of the 1P state, including mass and residue, are calculated based on the 1S state $+$ 1P state $+$ continuum scheme. In the second step, the continuum threshold, $s_0$, is adjusted to introduce the  2S state into the continuum.  It is important to mention that the mass and residue of the 1S state are considered as input quantities at this stage. Finally, in the third step, the previously described method is utilized for the next radial excitation, that is, the 2S state. In other words, the 1S state $+$ 1P state $+$ 2S state $+ $ continuum scheme is now employed to extract the physical parameters, mass and residue, for the 2S state.   For this purpose, the value of $s_0$  is increased again, and the masses and residues of the 1S and 1P states are used as inputs. In the next section, a complete numerical analysis of the results for these states will be presented. 

\section { Numerical Analyses }\label{sec:three}

The QCD sum rules derived for the masses and residues of triply heavy baryons with spin-3/2 in the ground, first orbital  and first radial excited states are  analyzed numerically in the current section using the required input parameters. A range of essential input parameters is required, including the quark masses in both the $\mathrm{\overline{MS}}$ and pole schemes, as well as the two-gluon and three-gluon condensates, which are presented in Table~\ref{tab:Parameter}. It is noteworthy that nonperturbative operators with mass dimensions higher than six are expressed as combinations of lower-mass-dimension operators, based on the hypothesis of QCD factorization. In the previous section, we utilized this  hypothesis to derive the expressions related to the heavy quark propagators. By incorporating these heavy quark propagators into Eq.~(\ref{QCDSide1}), higher-dimensional operators appear as products of lower-dimensional operators, whose numerical values are derived from the data provided in Table~\ref{tab:Parameter}.

As well as the parameters presented in Table~\ref{tab:Parameter}, the QCD sum rules for the mass and residue of triply heavy baryons with spin-3/2 include two auxiliary quantities: the Borel parameter, $M^2$,  and the continuum threshold parameter, $s_0 $, both of which must be determined. To specify these parameters, the QCD sum rules are subject to standard criteria of the method,  for instance, the weak dependence of  physical quantities on these auxiliary parameters, the convergence of OPE, and pole dominance.  The convergence of OPE is achieved when the contribution from the perturbative part dominates over the nonperturbative parts, and the higher the dimension of a nonperturbative operator, the smaller its contribution becomes.  Pole dominance ensures that the resonances under study provide a dominant contribution compared to higher excited and continuum states.  

\begin{table}[htb]
		\begin{tabular}{|c|c|}
			\hline\hline
Quantities&Measurements \\ \hline\hline
			$\bar{m}_{c}(\bar{m}_{c})$                                     & $1.27{}^{+0.02}_{-0.02}~\mathrm{GeV}$ \cite{ParticleDataGroup:2022pth}\\ \hline
						$m_{c}$                                     & $1.67{}^{+0.07}_{-0.07}~\mathrm{GeV}$ \cite{ParticleDataGroup:2022pth}\\  \hline
			$\bar{m}_{b}(\bar{m}_{b})$                                     & $4.18^{+0.03}_{-0.02}~\mathrm{GeV}$ \cite{ParticleDataGroup:2022pth}\\ \hline
						$m_{b}$                                     & $4.78{}^{+0.06}_{-0.06}~\mathrm{GeV}$ \cite{ParticleDataGroup:2022pth}\\ \hline
					$\langle \frac{\alpha_s}{\pi} G^2 \rangle $ & $0.012{}^{+0.004}_{-0.004}$ $~\mathrm{GeV}^4 $\cite{Belyaev:1982cd}\\ \hline
			$\langle g_s^3 G^3 \rangle $                & $ 0.57{}^{+0.29}_{-0.29}$ $~\mathrm{GeV}^6 $\cite{Narison:2015nxh}\\ 
			\hline\hline
		\end{tabular}
		\caption{List of input parameters used in the analysis.}
		\label{tab:Parameter}
	\end{table}

Therefore, the upper limit of the $M^2$ parameter is determined from the condition that the contributions from the first three resonances (FTRs)  surpass those from higher excited states and the continuum. Within the standard framework, the following conditions must be met:
\begin{eqnarray}
\mathrm{FTRC}=\frac{\Pi(s_0,M^2)}{\Pi(\infty,M^2)}\geq 0.5.
\end{eqnarray}
The lower limit for the $M^2$  is set by the requirement of OPE series convergence. This means that the perturbative contribution must dominate over the nonperturbative terms, and nonperturbative operators of higher dimensions should have lower contributions. To formulate this requirement, we define the following ratio:
\begin{equation}
	 \frac{\Pi ^{\mathrm{Dim8}}(s_0,M^2)}{\Pi (s_0,M^2)}\le\ 0.05.
	  	\label{eq:Convergence}
	  \end{equation} 

The last auxiliary parameter that needs to be determined is the continuum threshold, $ s_0 $. Its value is limited by the energy of the next excited state and is state-dependent, meaning it is different for the ground state, the first orbital and the first radial excitation. To limit the contribution of higher excited states in the calculations, $s_0$ is chosen based on the specific state under consideration. Table~\ref{results} lists the valid ranges  for the $M^2$ and the  $s_0$ across all channels and in both schemes.

\begin{figure}[h!]
	\begin{center}
		\includegraphics[width=.43\textwidth]{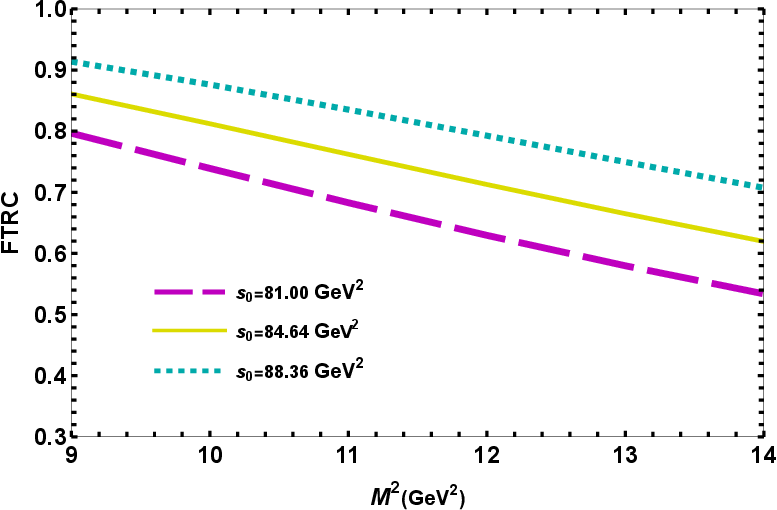}
	\end{center}
	\caption{ Dependence of FTRC on $M^2$ for three distinct values of $s_0$ in the $ \overline{\Omega}^*_{ccb}$  state. }
	\label{FTRC}
\end{figure}

The examination of  the first three resonances' contribution    (FTRC)  convergence provides valuable insights. Hence, we analyze the behavior of the FTRC within specified working intervals of auxiliary parameters.  Fig.~ \ref{FTRC} illustrates the dependence of the FTRC on $M^2$  for three fixed values of  $s_0$  for the particle  $\overline{\Omega}^*_{ccb}$  as an example. The results confirm that within the  valid renges of auxiliary parameters, the dominance of the pole contribution for the considered resonances is well maintained. In these calculations, the contribution of six-dimensional operators does not exceed 5\% in average values of the auxiliary parameters, and the contributions of higher-dimensional operators (eight)  remain below 1\%.  As a result, the requirements of the method are fully satisfied.

\begin{figure}[h!]
	\begin{center}
		\includegraphics[totalheight=4.5cm,width=6.5cm]{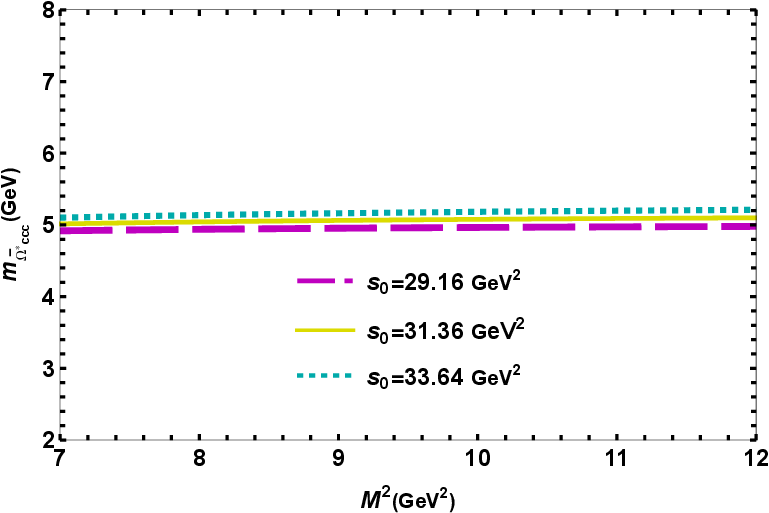}
		\includegraphics[totalheight=4.5cm,width=6.5cm]{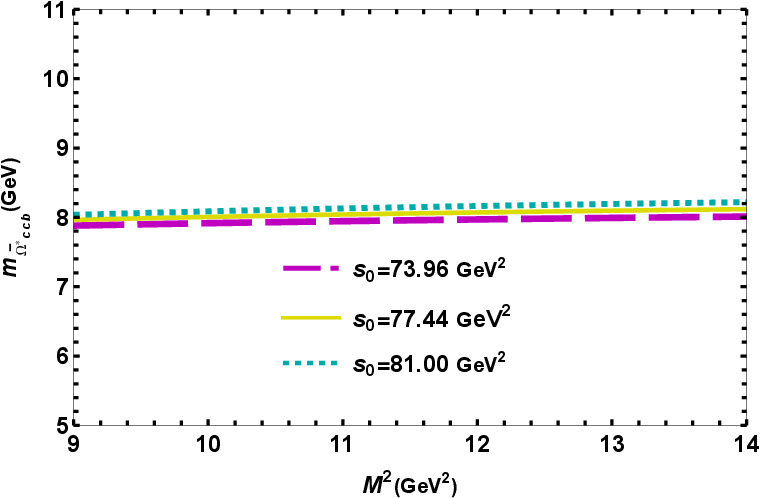}
		\includegraphics[totalheight=4.5cm,width=6.5cm]{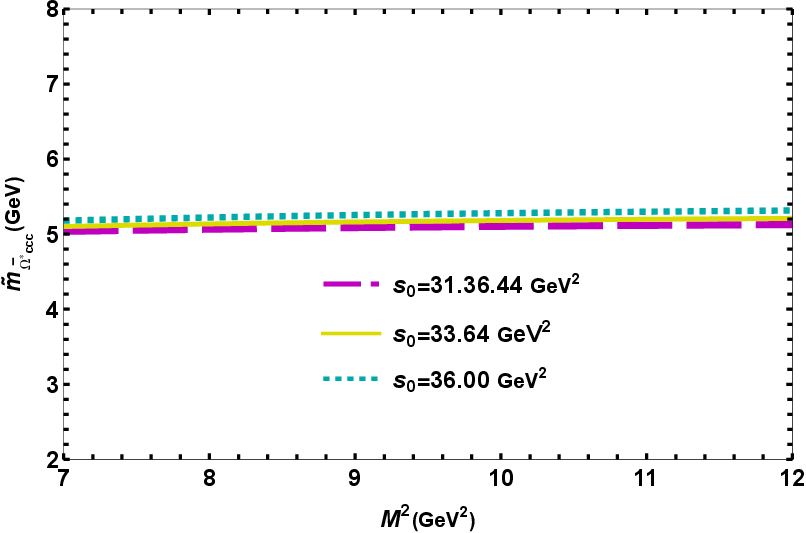}
		\includegraphics[totalheight=4.5cm,width=6.5cm]{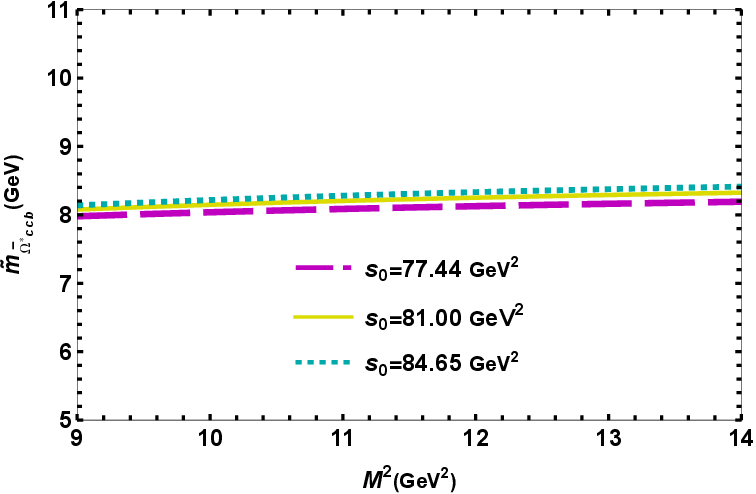}
		\includegraphics[totalheight=4.5cm,width=6.5cm]{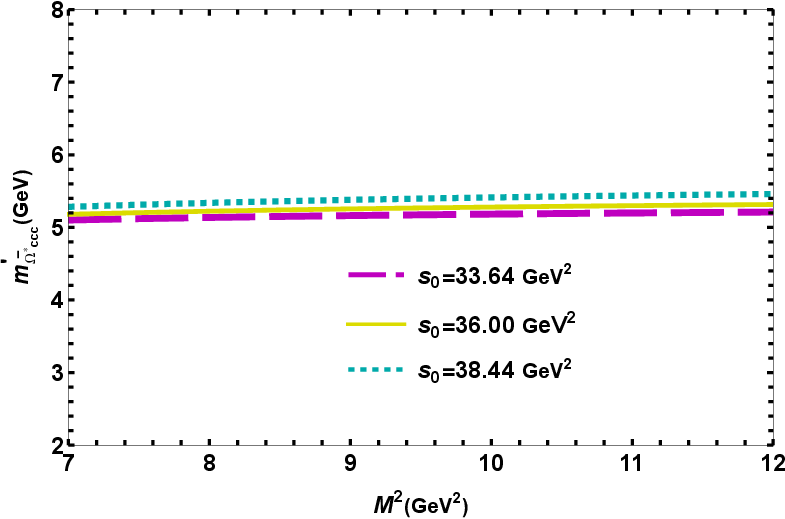}
		\includegraphics[totalheight=4.5cm,width=6.5cm]{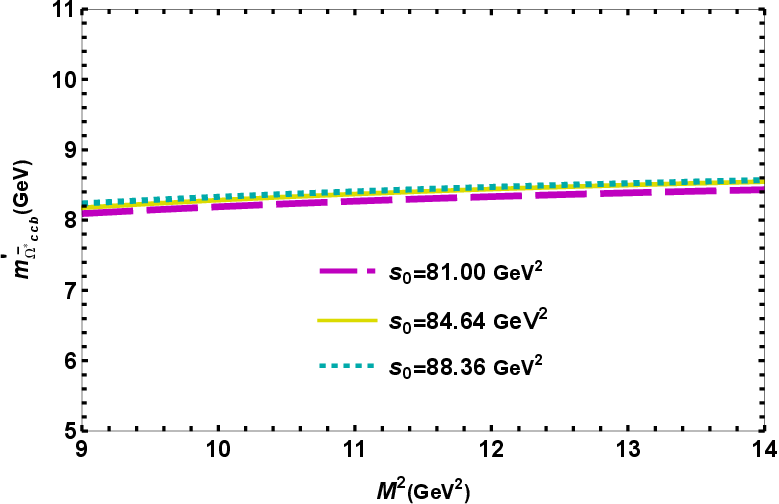}
		\end{center}
		\caption{\textbf{Left column:} Mass dependence of $\overline{\Omega}^*_{ccc}$ for the first $(1S)$, second $(1P)$ and third $(2S)$ resonances with respect to $M^2$ for different values of  $s_0$. \textbf{Right column:} Mass dependence of $	\overline{\Omega}^*_{ccb}$  for the first $(1S)$, second $(1P)$ and third $(2S)$ resonances with respect to $M^2$ for different values of  $s_0$..}
\label{gr}
\end{figure}

\begin{figure}[h!]
	\begin{center}
		\includegraphics[totalheight=4.5cm,width=6.5cm]{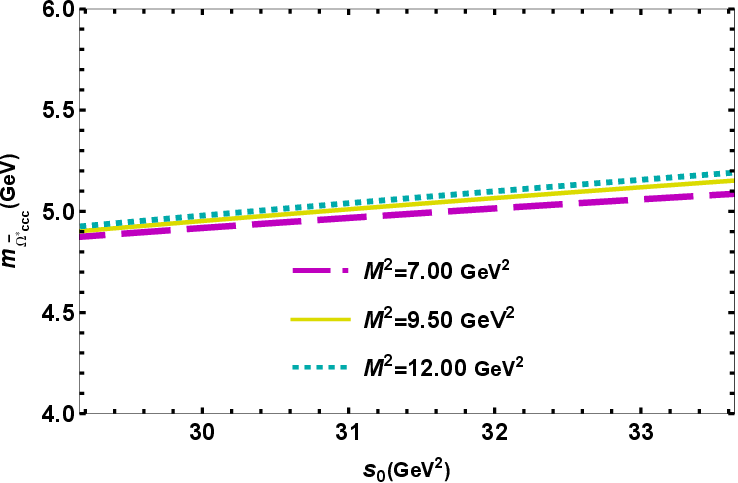}
		\includegraphics[totalheight=4.5cm,width=6.5cm]{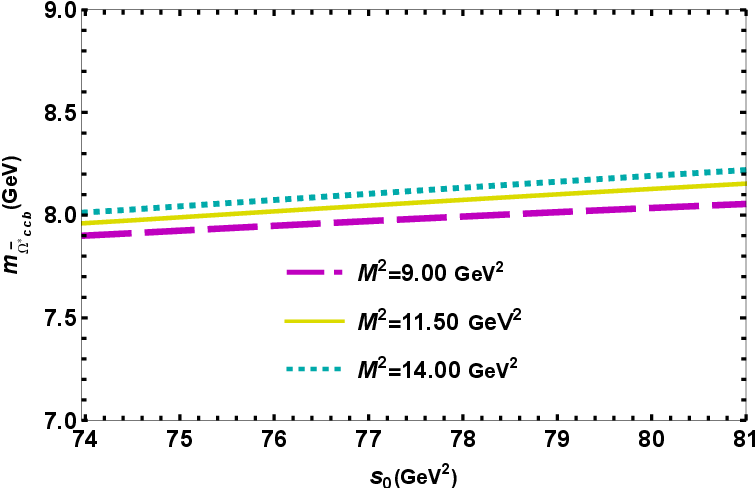}
		\includegraphics[totalheight=4.5cm,width=6.5cm]{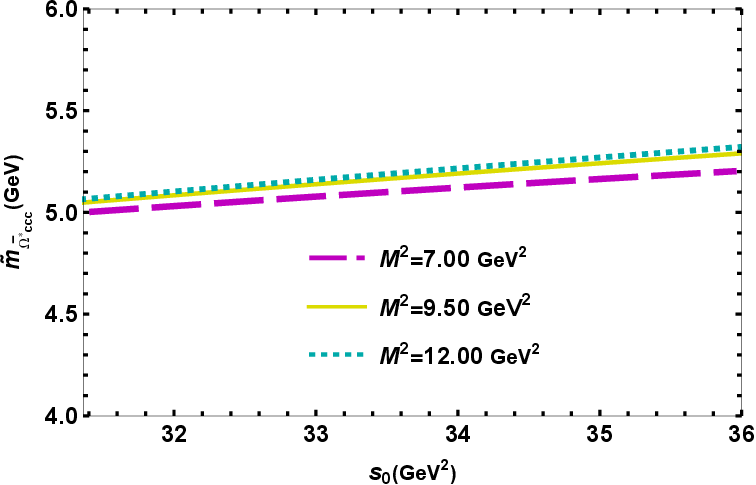}
		\includegraphics[totalheight=4.5cm,width=6.5cm]{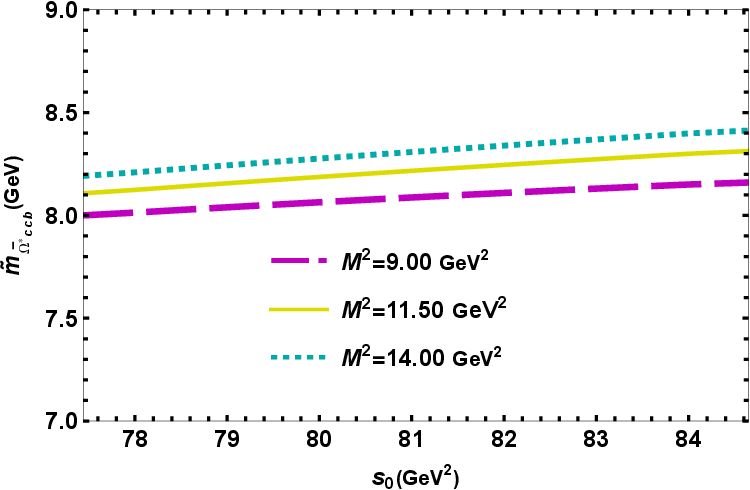}
		\includegraphics[totalheight=4.5cm,width=6.5cm]{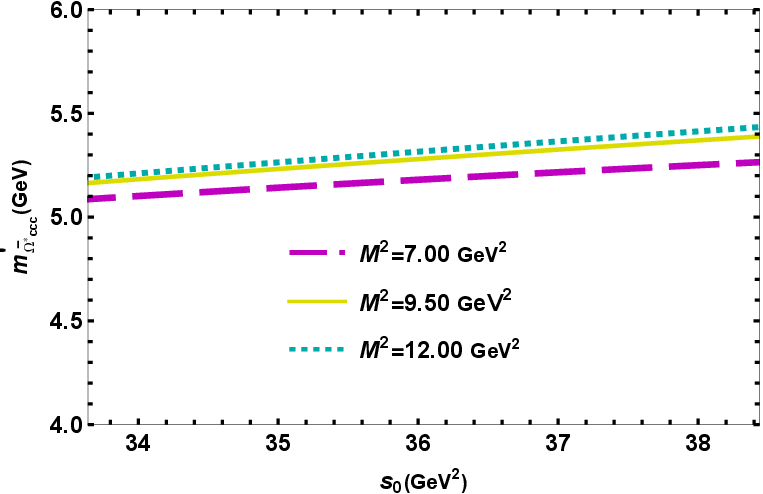}
		\includegraphics[totalheight=4.5cm,width=6.5cm]{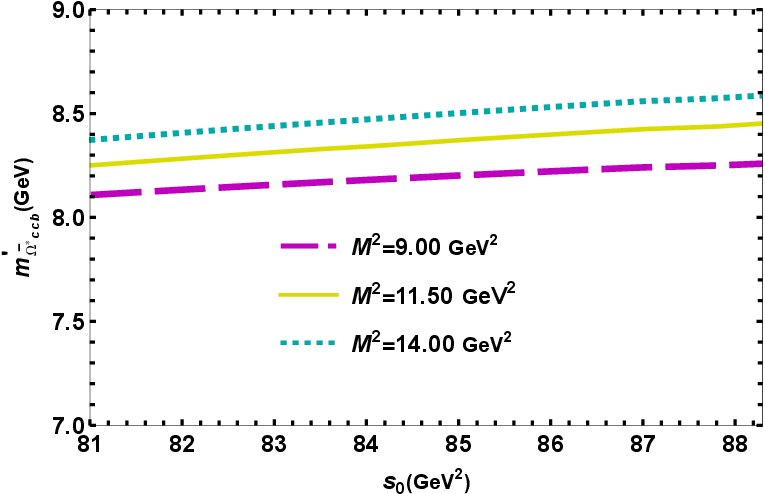}
					\end{center}
	\caption{\textbf{Left column:}  Mass dependence of $	\overline{\Omega}^*_{ccc}$  for the first $(1S)$, second $(1P)$ and third $(2S)$ resonances with respect to $s_0$ for different values of $M^2$. \textbf{Right column:} Mass dependence of $\overline{	\Omega}^*_{ccb}$ for the first $(1S)$, second $(1P)$ and third $(2S)$ resonances with respect to $s_0$ for different values of $M^2$.}
	\label{gr3}
\end{figure}

Based on the analyses conducted, we noticed that the physical quantities exhibit a weak dependence to the auxiliary parameters within the defined range of   $M^2$  and  $s_0$. To demonstrate the stability and impact of auxiliary parameters on the predicted masses, we have plotted  Figs.   \ref{gr} and \ref{gr3}. The results indicate remarkable stability of the masses concerning the $M^2$ and $s_0$ within the relevant working ranges.  It is noteworthy that our findings indicate  minimal sensitivity to auxiliary parameters for all triply heavy  baryons with spin-3/2. For this reason, we have included only the plots corresponding to the $ \overline{\Omega}^*_{ccc}$ and $ \overline{\Omega}^*_{ccb}$ states as examples.

After determining the relevant ranges for the auxiliary parameters, we have presented  the computed values  of the masses and residues in the two pole and $\mathrm{\overline{MS}}$ schemes for the studied states in Table~\ref{results}.  As previously mentioned, our method involves a step-by-step advancement in Eq.~ (\ref{Eq:cor:match1}), which leads to the calculation of the ground , the first orbital excitation and the first radial excitation states.  As discussed earlier, initially, by selecting an appropriate range for the threshold parameter, as also indicated in Table~\ref{results}, we compute the mass and residue of the triply heavy baryons with spin-3/2  in the 1S state within the framework of the 1S state $+$ continuum scheme. In the second step, by increasing the threshold parameter to define an appropriate range and using the computed values for the 1S state as input, we calculate the mass and residue of the second resonance (1P) state through the 1S state $+ $ 1P state $+$ continuum scheme. The results of the 1P state are also presented in Table~\ref{results}. Finally, by adjusting the range of the threshold parameter and utilizing the computed values from the previous two stages as input parameters, we determine the mass and residue of the 2S state, which corresponds to the third resonance, using the 1S state $+$ 1P state $+$ 2S  state $+$ continuum scheme. It should be emphasized that the obtained masses and residues of two Lorentz structures have approximately equivalent values. Hence, we solely consider the results derived from   the $\not\!q g_{\mu\nu}$ structure in our analysis. The uncertainties related to the auxiliary parameters, along with systematic errors from additional input parameters, are principal factors contributing to the uncertainty in our numerical results.  It should be noted that the magnitude of these uncertainties in mass values is significantly less than that in residues. This is because mass is derived from the ratio of two sum rules, which cancel each other's errors, while residues are derived from a single sum rule, as is clearly observed from Eqs.~(\ref{Eq:mass:Groundstates1}) and (\ref{Eq:residumass:Groundstates1}). 
 \begin{table}[!h]
	\begin{tabular}{|c|c|c|c|c|c|}
			\hline\hline
		Particle  & State &$M^2~(\mathrm{GeV^2})$&$\sqrt{s_0}(GeV)$  & m $(\mathrm{GeV})$ & $\lambda~(\mathrm{GeV^3})$ \\ \hline\hline
		\multirow{3}{*}{} &$\overline{\Omega}_{ccc}^*(\frac{3}{2}^+)(1S)$ &$7.00-12.00$&$ 5.60\pm0.20$&$ 5.06{}^{+0.14}_{-0.14}$& $0.38{}^{+0.06}_{-0.06} $ \\ \cline{2-6} 
		$\overline{\Omega}_{ccc}^*$    &$\overline{\Omega}^*_{ccc}(\frac{3}{2}^-)(1P)$ &$7.00-12.00$& $ 5.80\pm0.20$& $5.17{}^{+0.14}_{-0.13}$& $0.44{}^{+0.06}_{-0.07}$\\ \cline{2-6} 
		&$\overline{\Omega}_{ccc}^*(\frac{3}{2}^+)(2S)$ &$7.00-12.00$& $ 6.00\pm0.20$& $5.26{}^{+0.19}_{-0.16}$& $0.50{}^{+0.11}_{-0.10}$  \\ \hline\hline
		\multirow{3}{*}{} &$\Omega_{ccc}^*(\frac{3}{2}^+)(1S)$ &$7.00-12.00$&$ 5.60\pm0.20$&$ 5.20{}^{+0.30}_{-0.45}$& $0.25{}^{+0.12}_{-0.13} $ \\ \cline{2-6} 
		$\Omega_{ccc}^*$    &$\Omega_{ccc}^*(\frac{3}{2}^-)(1P)$ &$7.00-12.00$& $ 5.80\pm0.20$& $5.56{}^{+0.12}_{-0.14}$& $0.38{}^{+0.13}_{-0.14}$\\ \cline{2-6} 
		&$\Omega_{ccc}^*(\frac{3}{2}^+)(2S)$ &$7.00-12.00$& $ 6.00\pm0.20$& $5.70{}^{+0.16}_{-0.15}$& $0.54{}^{+0.22}_{-0.22}$  \\ \hline\hline
		\multirow{3}{*}{} &$\overline{\Omega}^*_{ccb}(\frac{3}{2}^+)(1S)$ &$9.00-14.00$&$ 8.80\pm0.20$&$ 8.06{}^{+0.16}_{-0.17}$& $0.36{}^{+0.07}_{-0.08} $ \\ \cline{2-6} 
		$\overline{\Omega}_{ccb}^*$    &$\overline{\Omega}^*_{ccb}(\frac{3}{2}^-)(1P)$ &$9.00-14.00$& $ 9.00\pm0.20$& $8.21{}^{+0.20}_{-0.23}$& $0.40{}^{+0.04}_{-0.10}$\\ \cline{2-6} 
		&$\overline{\Omega}_{ccb}^*(\frac{3}{2}^+)(2S)$ &$9.00-14.00$& $ 9.20\pm0.20$& $8.39{}^{+0.25}_{-0.30}$& $0.44{}^{+0.03}_{-0.09}$  
		\\ \hline\hline
		\multirow{3}{*}{} &$\Omega_{ccb}^*(\frac{3}{2}^+)(1S)$ &$9.00-14.00$&$ 8.80\pm0.20$&$ 8.58{}^{+0.14}_{-0.17}$& $0.15{}^{+0.03}_{-0.03} $ \\ \cline{2-6} 
		$\Omega_{ccb}^*$    &$\Omega_{ccb}^*(\frac{3}{2}^-)(1P)$ &$9.00-14.00$& $ 9.00\pm0.20$& $8.89{}^{+0.12}_{-0.16}$& $0.17{}^{+0.01}_{-0.07}$\\ \cline{2-6} 
		&$\Omega_{ccb}^*(\frac{3}{2}^+)(2S)$ &$9.00-14.00$& $ 9.20\pm0.20$& $9.10{}^{+0.13}_{-0.14}$& $0.44{}^{+0.20}_{-0.26}$  \\ \hline\hline
		\multirow{3}{*}{} &$\overline{\Omega}^*_{bbc}(\frac{3}{2}^+)(1S)$ &$12.00-16.00$&$ 12.00\pm0.20$&$ 11.02{}^{+0.13}_{-0.12}$& $0.13{}^{+0.01}_{-0.01} $ \\ \cline{2-6} 
		$\overline{\Omega}_{bbc}^*$    &$\overline{\Omega}_{bbc}^*(\frac{3}{2}^-)(1P)$ &$12.00-16.00$& $ 12.20\pm0.20$& $11.14{}^{+0.13}_{-0.13}$& $0.15{}^{+0.02}_{-0.01}$\\ \cline{2-6} 
		&$\overline{\Omega}_{bbc}^*(\frac{3}{2}^+)(2S)$ &$12.00-16.00$& $ 12.40\pm0.20$& $11.23{}^{+0.15}_{-0.14}$& $0.17{}^{+0.02}_{-0.02}$  
		\\ \hline\hline
		\multirow{3}{*}{} &$\Omega_{bbc}^*(\frac{3}{2}^+)(1S)$ &$12.00-16.00$&$ 12.00\pm0.20$&$ 11.74{}^{+0.13}_{-0.13}$& $0.11{}^{+0.04}_{-0.03} $ \\ \cline{2-6} 
		$\Omega_{bbc}^*$    &$\Omega_{bbc}^*(\frac{3}{2}^-)(1P)$ &$12.00-16.00$& $ 12.20\pm0.20$& $11.96{}^{+0.10}_{-0.10}$& $0.16{}^{+0.03}_{-0.05}$\\ \cline{2-6} 
		&$\Omega_{bbc}^*(\frac{3}{2}^+)(2S)$ &$12.00-16.00$& $ 12.40\pm0.20$& $12.14{}^{+0.11}_{-0.10}$& $0.20{}^{+0.05}_{-0.06}$  
		\\ \hline\hline
		\multirow{3}{*}{} &$\overline{\Omega}^*_{bbb}(\frac{3}{2}^+)(1S)$ &$14.00-18.00$&$ 15.30\pm0.20$&$ 13.97{}^{+0.13}_{-0.14}$& $0.22{}^{+0.02}_{-0.02} $ \\ \cline{2-6} 
		$\overline{\Omega}_{bbb}^*$    &$\overline{\Omega}^*_{bbb}(\frac{3}{2}^-)(1P)$ &$14.00-18.00$& $ 15.50\pm0.20$& $14.10{}^{+0.13}_{-0.13}$& $0.24{}^{+0.02}_{-0.02}$\\ \cline{2-6} 
		&$\overline{\Omega}_{bbb}^*(\frac{3}{2}^+)(2S)$ &$14.00-18.00$& $ 15.70\pm0.20$& $14.20{}^{+0.13}_{-0.12}$& $0.26{}^{+0.02}_{-0.02}$  
		\\ \hline\hline
		\multirow{3}{*}{} &$\Omega_{bbb}^*(\frac{3}{2}^+)(1S)$ &$14.00-18.00$&$ 15.30\pm0.20$&$ 14.96{}^{+0.13}_{-0.14}$& $0.10{}^{+0.02}_{-0.02} $ \\ \cline{2-6} 
		$\Omega_{bbb}^*$    &$\Omega_{bbb}^*(\frac{3}{2}^-)(1P)$ &$14.00-18.00$& $ 15.50\pm0.20$& $15.15{}^{+0.10}_{-0.12}$ & $0.24{}^{+0.02}_{-0.02}$\\ \cline{2-6} 
		&$\Omega_{bbb}^*(\frac{3}{2}^+)(2S)$ &$14.00-18.00$& $ 15.70\pm0.20$& $15.44{}^{+0.07}_{-0.05}$& $0.35{}^{+0.06}_{-0.05}$  
		\\ \hline\hline								 	
			\end{tabular}
	\caption{The stability intervals of the auxiliary parameters along with the calculated mass and residue results. The triply heavy  baryons  denoted with an overline correspond to the $\mathrm{\overline{MS}}$ scheme for heavy quarks, whereas those without an overline are associated with the pole scheme.}  
		\label{results}
\end{table}
	\begin{table}[h!]
	\scalebox{0.78}{\begin{tabular}{|c|c|c|c|c|c|c|c|c|c|c|c|c|}
			\hline\hline
	\multicolumn{1}{|c|}{particle}&  \multicolumn{3}{c|}{$\overline{\Omega}^*_{ccc}$} &  \multicolumn{3}{c|}{$\overline{\Omega}^*_{ccb}$}&  \multicolumn{3}{c|}{$\overline{\Omega}^*_{bbc}$} &  \multicolumn{3}{c|}{$\overline{\Omega}^*_{bbb}$} \\ \hline
	resonant state&  $1S$ & $1P$ & $2S$ & $1S$ & $1P$ & $2S$ & $1S$ & $1P$ & $2S$& $1S$ & $1P$ & $2S$\\ \hline\hline
	Present\,Work&$ 5.06{}^{+0.14}_{-0.14}$&$5.17{}^{+0.14}_{-0.13}$&$5.26{}^{+0.19}_{-0.16}$&$ 8.06{}^{+0.16}_{-0.17}$&$8.21{}^{+0.20}_{-0.23}$&$8.39{}^{+0.25}_{-0.30}$&$ 11.02{}^{+0.13}_{-0.12}$&$11.14{}^{+0.13}_{-0.13}$&$11.23{}^{+0.15}_{-0.14}$&$ 13.97{}^{+0.13}_{-0.14}$&$14.10{}^{+0.13}_{-0.13}$&$14.20{}^{+0.13}_{-0.12}$\\ \hline   
	 Non-relativistic quark model \cite{Roberts:2007ni}&$4.96$&$5.16$&$5.31$&$8.26$&$8.42$&$8.55$&$11.55$&$11.71$&$11.79$&$14.83$&$14.97$&$15.08$\\ 	\hline   
	 Non-relativistic quark model \cite{Shah:2018div}&$-$&$-$&$-$&$-$&$-$&$-$&$11.29$&$11.56$&$11.77$&$-$&$-$&$-$\\ 	\hline
	 Non-relativistic quark model \cite{Shah:2018bnr}&$-$&$-$&$-$&$8.04$&$8.37$&$8.62$&$-$&$-$&$-$&$-$&$-$&$-$\\ \hline
	 Non-relativistic quark model \cite{Ortiz-Pacheco:2023kjn}&$4.90$&$5.07$&$-$&$-$&$-$&$-$&$-$&$-$&$-$&$14.86$&$14.96$&$-$\\ \hline 	 
	Relativistic quark model \cite{Yang:2019lsg}&$4.79$&$5.12$&$5.58$&$8.02$&$8.30$&$8.46$&$11.22$&$11.48$&$11.62$&$14.39$&$14.68$&$14.80$\\ \hline
	           Relativistic quark model \cite{Faustov:2021qqf}&$4.71$ &$5.02$&$5.13$&$7.99$ &$8.26$&$8.36$&$11.21$ &$11.42$&$11.51$&$14.46$ &$14.70$&$14.81$		\\ \hline
	                      Faddeev equation approach \cite{Qin:2019hgk}&$4.76$&$5.02$&$5.15$&$7.96$&$8.27$&$8.42$&$11.16$&$11.52$&$11.70$&$14.37$&$14.77$&$14.98$\\ \hline
             Sum rule method \cite{Wang:2011ae}&$4.76{}^{+0.14}_{-0.14}$  &$4.88{}^{+0.15}_{-0.15}$&$-$&$7.60{}^{+0.13}_{-0.13}$  &$7.73{}^{+0.13}_{-0.13}$&$-$&$10.46{}^{+0.12}_{-0.12}$  &$10.59{}^{+0.11}_{-0.11}$&$-$&$13.40{}^{+0.10}_{-0.10}$  &$13.52{}^{+0.11}_{-0.11}$&$-$\\ \hline
               Sum rule method \cite{Aliev:2014lxa}&$4.72{}^{+0.12}_{-0.12}$	& $4.9{}^{+0.1}_{-0.1}$&$-$&$8.07{}^{+0.10}_{-0.10}$	& $8.35{}^{+0.10}_{-0.10}$&$-$&$11.35{}^{+0.15}_{-0.15}$	& $11.5{}^{+0.2}_{-0.2}$&$-$&$14.3{}^{+0.2}_{-0.2}$	& $14.9{}^{+0.2}_{-0.2}$&$-$\\ \hline
               	Sum rule method \cite{Wang:2020avt}&$4.81{}^{+0.10}_{-0.10}$  &$-$&$-$&$8.03{}^{+0.08}_{-0.08}$  &$-$&$-$&$11.22{}^{+0.08}_{-0.08}$  &$-$&$-$&$14.43{}^{+0.09}_{-0.09}$  &$-$&$-$\\ \hline  
	        Regge trajectory model \cite{Wei:2015gsa} &$4.83{}^{+0.08}_{-0.08}$&$5.07{}^{+0.10}_{-0.10}$&$-$&$-$&$-$&$-$&$-$&$-$&$-$&$-$&$-$&$-$\\ \hline  
Regge trajectory model \cite{Oudichhya:2023pkg}&$-$&$-$&$-$&$8.22$  &$8.50$&$8.63$&$11.54$  &$11.88$&$11.77$&$-$&$-$&$-$\\ \hline
  Effective Hamiltonian model \cite{Serafin:2018aih}&$4.79$&$5.10$&$5.30$&$8.30$&$8.49$&$8.60$&$11.21$&$11.43$&$11.58$&$14.34$&$14.64$&$14.83$\\ \hline    
	 		\hline
	\end{tabular} }
	\caption{The mass values of the first three resonant states (expressed in $\mathrm{GeV}$)  for the baryons $\overline{\Omega}_{ccc}^*$ , $\overline{\Omega}_{ccb}^*$ , $\overline{\Omega}_{bbc}^*$ and  $\overline{\Omega}_{bbb}^*$,  and their comparison with theoretical predictions from other studies.}
		\label{comper results}
\end{table}
As presented in Table~\ref{results}, the mass difference between the  1S and 1P (2S) for the particles under investigation is as follows:	$\overline{\Omega}^*_{ccc}$: 0.11 (0.20) $\mathrm{GeV}$,	$\Omega^*_{ccc}$: 0.36 (0.50) $\mathrm{GeV}$, $\overline{\Omega}^*_{ccb}$: 0.15 (0.33) $\mathrm{GeV}$,	$\Omega^*_{ccb}$: 0.31(0.52) $\mathrm{GeV}$, $\overline{\Omega}^*_{bbc}$: 0.12 (0.21) $\mathrm{GeV}$, $\Omega^*_{bbc}$: 0.22 (0.40) $\mathrm{GeV}$, $\overline{\Omega}^*_{bbb}$: 0.13 (0.23) $\mathrm{GeV}$, $\Omega^*_{bbb}$: 0.19 (0.48) $\mathrm{GeV}$( the particles denoted with an overline correspond to the $\mathrm{\overline{MS}}$ scheme for heavy quarks, whereas those without an overline are associated with the pole scheme).

 To provide a comprehensive comparison and in light of the lack of available measurements for the triply heavy baryons, Table~\ref{comper results} presents predictions obtained from various theoretical approaches for their masses.  As can be illustrated  in Table~\ref{comper results}, some of them only include the masses of the ground state, while others report excited states also. It is worth mentioning that most of these studies do not specify quark masses within a particular scheme; therefore, we compare their predicted masses and residues with our findings derived from the $\mathrm{\overline{MS}}$ scheme for the masses of heavy quarks (b and c).    We perform these calculations by including nonperturbative operators up to eight mass dimensions to enhance accuracy. 
In previous work, calculations were carried out up to four mass dimensions with a smaller set of resonances using the QCD sum rules \cite{Aliev:2014lxa}.	As observed from the data presented in Table~\ref{comper results},   and crucially considering the $ \pm 0.14 \, \text{GeV}$ calculated uncertainty for the $\overline{\Omega}^*_{ccc}$ (1S) state, its masses obtained  using  the non-relativistic quark model \cite{Roberts:2007ni,Ortiz-Pacheco:2023kjn}, are almost in alignment with our predictions within the uncertainty range for the FTRs. The predictions made by the  relativistic quark model \cite{Yang:2019lsg,Faustov:2021qqf} for the mentioned particle exhibit slight differences from our results in the 1S state and the 2S state, while they are fully in agreement within the uncertainty range for the first  orbital excited state (1P). This comparison for  $\overline{\Omega}^*_{ccc}$ using the Faddeev equation method \cite{Qin:2019hgk} shows some differences in the 1S state and the 1P  state, where the difference for the 1P state is very small and also aligns well with the first radial excited state (2S). The prediction of the mass of $\overline{\Omega}^*_{ccc}$ in its ground state by the Regge trajectories  \cite{Wei:2015gsa} and the effective Hamiltonian method \cite{Serafin:2018aih} shows slight differences from our results within the uncertainty range. However, for the first orbital excited state, the results are fully consistent. Our results for the ground state of   $\overline{\Omega}^*_{ccc}$  exhibit small differences when compared to predictions made using   the sum rule method \cite{Wang:2011ae,Aliev:2014lxa, Wang:2020avt} while it is in good consistency with the mass prediction for the first orbital excited state \cite{Wang:2011ae,Aliev:2014lxa}. Our obtained masses for  the FTRs of $\overline{\Omega}^*_{ccb}$ agree well with the theoretical estimates from different frameworks such as the non- relativistic quark  model approach \cite{Shah:2018bnr}, the relativistic quark model \cite{Yang:2019lsg, Faustov:2021qqf}, the Faddeev equation \cite{Qin:2019hgk} and the QCD sum rule method \cite{Aliev:2014lxa,Wang:2020avt} within the indicated uncertainties.  Our findings regarding $\overline{\Omega}^*_{ccb}$ show a slight difference from predictions from  approaches  like the non- relativistic quark \cite{Roberts:2007ni}, the QCD sum rules \cite{Wang:2011ae}, Regge trajectories  \cite{Oudichhya:2023pkg} and the effective Hamiltonian method \cite{Serafin:2018aih}  for the  1S and 1P states, while our obtained mass of the 2S state is in excellent agreement with the mentioned theoretical predictions in Refs. \cite{Wei:2015gsa,Serafin:2018aih}, within errors. The masses calculated for $\overline{\Omega}^*_{bbc}$ across the FTRs show discrepancies relative to the predictions of the  mentioned theoretical methods within the uncertainty range as follows:  for the predictions of   the non-relativistic method \cite{Roberts:2007ni} (\cite{Shah:2018div}), there is a discrepancy of approximately 3\% (1\%) in the 1S state, 3\%(2\%) in the 1P state and 3\% (3\%) in  the 2S state.  According to the mass predictions in the  relativistic approach  \cite{Yang:2019lsg} (\cite{Faustov:2021qqf}), discrepancies of  0.6\% (0.5\%), 1\% (1\%), and 2\% (1\%) are observed in the 1S , 1P and 2S states, respectively. The predicted  results using the Faddeev equation \cite{Qin:2019hgk} illustrate a discrepancy of 0.08\% in the ground state and a 2\% discrepancy in both subsequent resonances. Results computed using  Regge trajectories  \cite{Oudichhya:2023pkg} show differences of 3\%, 5\%  and 3\% for the  1S, 1P and 2S states, respectively, while this order for the predictions of the effective Hamiltonian method \cite{Serafin:2018aih} is 0.5\%, 1\%, and 1\%. The differences in mass predictions in the sum rules method \cite{Aliev:2014lxa} (\cite{Wang:2011ae}) are of 2\% (0.4\%) in the ground state and  2\% (0.2\%) in the next calculated resonance (1P). The predicted results for the ground state by the sum rules in Ref.  \cite{Wang:2020avt} show a difference of about 0.4\% compared to our findings. The mass of $\overline{\Omega}^*_{bbb}$ in its ground state shows a discrepancy of 5\% compared to predicted values in  Refs.  \cite{Roberts:2007ni,Ortiz-Pacheco:2023kjn}. In the 1S state, results obtained in Refs. \cite{Faustov:2021qqf,Wang:2011ae}  exhibit a discrepancy of 2\% and calculations performed in Refs. \cite{Yang:2019lsg,Qin:2019hgk,Wang:2020avt,Serafin:2018aih} show a discrepancy of 1\% with our calculation. Our prediction in the 1S state  aligns with previous results in Ref. \cite{Aliev:2014lxa} within the uncertainty range. In the first orbital excitation of  $\overline{\Omega}^*_{bbb}$, our predicted mass shows a discrepancy of 5\% compared to results obtained in Refs.  \cite{Roberts:2007ni,Ortiz-Pacheco:2023kjn}, 3\% with predictions made in Refs. \cite{Yang:2019lsg,Faustov:2021qqf,Qin:2019hgk,Aliev:2014lxa} and  2\% with calculated masses in Refs\cite{Wang:2011ae,Serafin:2018aih}. For the first radial excited state of particle $\overline{\Omega}^*_{bbb}$, our results show discrepancies of 5\%, 4\% and 3\%, respectively, when compared to existing results in Ref. \cite{Roberts:2007ni}, Ref. \cite{Qin:2019hgk} and Refs.  \cite{Yang:2019lsg,Faustov:2021qqf,Serafin:2018aih}. 

Furthermore, as mentioned earlier, we have calculated the residues of the 1S, 1P and 2S states, in light of the effect of nonperturbative terms up to the eight mass dimension, and have presented them in Table~\ref{results}  along with the error interval. Conversely, only a limited number of studies on the residue quantities of triply heavy baryons with spin-3/2, which serve as input for analyzing various related decay processes, can be found in the literature. Table~\ref{comper results1} summarizes the available residue predictions with the aim of comparing them.  The residues of triply heavy baryons with spin-3/2 in the 1S and 1P states have been computed using the QCD sum rule technique \cite{Wang:2011ae,Aliev:2014lxa}. In the 1S and 1P states, overall we see a significant difference in the values of the presented residues and existing studies (there are some consistencies between our results and the prediction of residue in Ref. \cite{Wang:2011ae} for $\overline{\Omega}^*_{ccb}$ in the ground state,  the prediction of residue in Ref. \cite{Wang:2011ae} for $\overline{\Omega}^*_{ccb}$ in the first orbital excited state and the prediction of residue in Ref. \cite{Aliev:2014lxa} for $\overline{\Omega}^*_{bbb}$ in the first orbital excited state, within the indicated uncertainties). 

At the closing of this section, it is worth mentioning that our results can be used as input for investigating various decay channels of triply heavy spin-3/2 baryons and may also facilitate the search for these unseen particles for experimental groups.
\begin{table}[htb]
	\begin{tabular}{|c|c|c|c|c|}\hline\hline
	Particle  & State &Present\,Work   & Ref.\cite{Wang:2011ae}& Ref.\cite{Aliev:2014lxa} \\ \hline\hline
			\multirow{3}{*}{} &$\overline{\Omega}_{ccc}^*(\frac{3}{2}^+)(1S)$ & $0.38{}^{+0.06}_{-0.06} $  & $0.20{}^{+0.04}_{-0.04}$& $0.09{}^{+0.01}_{-0.01}$ \\ \cline{2-5} 
		$\overline{\Omega}_{ccc}^*$    &$\overline{\Omega}_{ccc}^*(\frac{1}{2}^-)(1P)$ & $0.44{}^{+0.06}_{-0.07}$ & $0.24{}^{+0.04}_{-0.04}$& $0.11{}^{+0.01}_{-0.01}$ \\ \cline{2-5} 
		&$\overline{\Omega}_{ccc}^*(\frac{3}{2}^+)(2S)$ & $0.50{}^{+0.11}_{-0.10}$ & $-$&$-$ \\ \hline\hline
		\multirow{3}{*}{} &$\overline{\Omega}_{ccb}^*(\frac{3}{2}^+)(1S)$ & $0.36{}^{+0.07}_{-0.08} $  & $0.26{}^{+0.05}_{-0.05}$& $0.06{}^{+0.01}_{-0.01}$ \\ \cline{2-5} 
		$\overline{\Omega}_{ccb}^*$    &$\overline{\Omega}_{ccb}^*(\frac{1}{2}^-)(1P)$ & $0.40{}^{+0.04}_{-0.10}$ & $0.32{}^{+0.06}_{-0.06}$& $0.07{}^{+0.01}_{-0.01}$ \\ \cline{2-5} 
		&$\overline{\Omega}_{ccb}^*(\frac{3}{2}^+)(2S)$ & $0.44{}^{+0.03}_{-0.09}$ & $-$&$-$ \\ \hline\hline		
		\multirow{3}{*}{} &$\overline{\Omega}_{bbc}^*(\frac{3}{2}^+)(1S)$ & $0.13{}^{+0.01}_{-0.01} $  & $0.39{}^{+0.09}_{-0.09}$& $0.08{}^{+0.01}_{-0.01}$ \\ \cline{2-5} 
		$\overline{\Omega}_{bbc}^*$    &$\overline{\Omega}_{bbc}^*(\frac{1}{2}^-)(1P)$ & $0.15{}^{+0.02}_{-0.01}$ & $0.47{}^{+0.10}_{-0.10}$& $0.09{}^{+0.01}_{-0.01}$ \\ \cline{2-5} 
		&$\overline{\Omega}_{bbc}^*(\frac{3}{2}^+)(2S)$ & $0.17{}^{+0.02}_{-0.02}$ & $-$&$-$ \\ \hline\hline	
		\multirow{3}{*}{} &$\overline{\Omega}_{bbb}^*(\frac{3}{2}^+)(1S)$ & $0.22{}^{+0.02}_{-0.02} $ & $0.66{}^{+0.15}_{-0.15}$& $0.14{}^{+0.02}_{-0.02}$ \\ \cline{2-5} 
		$\overline{\Omega}_{bbb}^*$    &$\overline{\Omega}_{bbb}^*(\frac{1}{2}^-)(1P)$ &  $0.24{}^{+0.02}_{-0.02}$ & $0.82{}^{+0.16}_{-0.16}$& $0.20{}^{+0.02}_{-0.02}$ \\ \cline{2-5} 
		&$\overline{\Omega}_{bbb}^*(\frac{3}{2}^+)(2S)$ & $0.26{}^{+0.02}_{-0.02}$ & $-$&$-$ \\ \hline\hline	
					\end{tabular}
\caption{The residue values of the first three resonant states (expressed in $\mathrm{GeV}$)  for the baryons $\overline{\Omega}_{ccc}^*$ , $\overline{\Omega}_{ccb}^*$ , $\overline{\Omega}_{bbc}^*$ and  $\overline{\Omega}_{bbb}^*$,  and their comparison with theoretical predictions from other studies.}
\label{comper results1}
\end{table}

\section { SUMMARY AND CONCLUSION }\label{sec:four}

Based on the quark model, the singly, doubly, and triply hadronic states containing heavy quarks are predicted. In this work, we investigated  the ground and excited states of the triply heavy baryons  with spin-3/2, considering positive and negative parity (the investigation of  these states with spin-1/2 was carried out in previous study \cite{Najjar:2024deh}).  The mass and residue are among the most fundamental parameters characterizing a particle and play a key role in the analysis of its interaction mechanisms and decay behaviors.	To analyze these states, we employ the sum rules, which serve as a predictive and reliable nonperturbative approach. To enhance the accuracy of mass and residue calculations for triply heavy baryons with spin-3/2 in the ground state (1S), the first orbital excitation (1P), and the first radial excitation (2S), we have incorporated nonperturbative contributions up to mass dimension eight. The masses and residues of the FTRs for the investigated particles have been determined by optimizing the auxiliary parameters specific to each state, as presented in Table~\ref{results}.  Moreover, in the absence of experimental data for the studied particles, we have exclusively compared our obtained masses and residues with the available theoretical predictions, as shown in Tables \ref{comper results} and \ref{comper results1}.	Our results for the masses show slight differences from some of the channels within the presented uncertainty range, with deviations ranging from 0.2 \% to 5\%.  In contrast, for the residues, we observe more significant discrepancies between our results and previous predictions.

For the interaction/decay of triply heavy baryons with/to other states, more accurate and detailed information regarding their mass and residue is essential. Therefore, our more accurate predictions could contribute to ongoing experimental searches for the missing members of baryons consisting of three heavy quarks or may be examined in future studies through related analyses. The identification of these particles in experiments will validate the success of the quark model in predicting particles with three heavy quarks, as well as represent another significant achievement in colliders.

\section*{ACKNOWLEDGMENTS} 

K. Azizi  is grateful  to Iran national science foundation (INSF) for the partial financial support provided under the elites Grant No. 4037888. He  also thanks   the CERN-Theory department for their support and warm hospitality.

\section*{APPENDIX: VARIOUS RELATIONS DERIVED FROM THE QCD SIDE OF THE ANALYSIS}

In this appendix, we provide the detailed expression for the individual components of the spectral density  $\rho^{pert}_{\not\!q ~g_{\mu\nu}}(s)$ as derived from our calculations: 
	\begin{eqnarray} 
		\rho^{pert}_{\not\!q ~g_{\mu\nu}}(s)&=\frac{\sqrt{3}}{4 \pi ^4}
		\int^{1}_{0} dz \int^{1-z}_{0} dr \ \frac{D\,\Theta(D)}{B^2\, C} \Big\{\- m_Q^2~A+2~m_Q m_{Q'}~z(\frac{A-B}{C})		\Big\}.
	\end{eqnarray} 
	Given the lengthy nature of the expressions for  $\Gamma^{dim-8}_{\not\!q g_{\mu\nu}}(M^2)$ and $\Gamma^{dim-6}_{\not\!q g_{\mu\nu}}(M^2)$, we provide the explicit form only for the coefficient of $\Gamma^{dim-4}_{\not\!q g_{\mu\nu}}(M^2)$:
		\begin{eqnarray} 
		\Gamma^{dim-4}_{\not\!q ~g_{\mu\nu}}(M^2)&=&\frac{1}{192 \pi ^2}
		\int^{1}_{0} dz \int^{1-z}_{0} dr \ \Big\langle\frac{\alpha_{s}GG}{\pi}\Big\rangle\ e^{-\frac{K}{M^2}}\Bigg\{A  ~C^7 m_Q^2 \Big[-C^4 \big[A^2 - 2 A ~B + B^2 - 3 A ~C r + B~ C r\big] \notag  \\
		& + &(-15 A + 7 B) C^4 r z 
		+ 
 6 (5 A - 3 B) C^3 r z^2 + 
 2 (-15 A + 11 B) C^2 r z^3+ (15 A - 11 B) C r z^4 \notag  \\
 & +& (A - 
    B)^2 \big[5 C^3 z - 10 C^2 z^2 + 10 C z^3 - 5 z^4\big]\Big] +\frac{m_Q^2 ~m_{Q'}^2}{24 \sqrt{3}  M^2 \pi^2}\Big[ \frac{(C^4 - 5 C^3 z + 10 C^2 z^2 - 10 C z^3 + 5 z^4)}{A^2~C~r^2}\notag  \\
    &+&\frac{r}{A~B}+\frac{2 (C - z) z(C^4 - 5 C^3 z + 10 C^2 z^2 - 10 C z^3 + 5 z^4)}{A^2~B~C~r^2}\notag  \\
    &+&\frac{(C - z)^2 z^2 (C^4 - 5 C^3 z + 10 C^2 z^2 - 10 C z^3 + 5 z^4)}{A^2~B^2~C~r^2} \Big]  +\frac{ 2  m_Q m_{Q'}}{z^4}\Big[-2 A^6 B - 3 A^5 B^2 + A^4 B^3 \notag\\
    &+& A^3 B^2 C r z (-1 + r + 3 z) + 
  A^2 z^5 (-5 C^4 z + 5 C z^4 - z^5 + C^3 (-3 r + 10 z^2) + 
     C^2 (1 - 10 z^3))\Big]\notag\\
     &+&\frac{m_Q ~m_{Q'}^3}{12 \sqrt{3}  M^2 \pi^2}\Big[\frac{A^2}{ C^3  r^2 z^4}\big(r^2 - 2 (-1 + z) z - r (1 + 2 z)\big)\notag\\
     &+& \frac{z}{B^2 C^3  r^2 } \big(5 C^4 z - 5 C z^4 + z^5 + C^3 (3 r - 10 z^2) + C^2 (-1 + 10 z^3)\big)\notag\\
     &+& \frac{z^2}{A^2 B~ C^3  r^2 } \big(-45 C^4 z^2 + 18 C z^5 - 3 z^6 + 
  C^2 z (18 + r (-54 + (54 - 17 r) r) - 45 z^3)\notag\\
  & + &
  C^3 (-3 + (-3 + r) r (-3 + 2 r) + 60 z^3)\big)\Big] +\frac{m_Q ^3~m_{Q'}}{12 \sqrt{3}  M^2 \pi^2}\Big[\frac{-   z^4}{A~ B~ C  ~ r^2}\notag\\
  &+&\frac{z  }{ B^2 C^2   r^3}\big(5 C^4 z - 5 C z^4 + z^5 + C^3 (3 r - 10 z^2) + C^2 (-1 + 10 z^3)\big)\notag\\
  &+&\frac{z  }{ A^3  C^2   r^3} \big(-C^6 r + C^5 (2 + 3 r) z + 2 C^4 (-6 + r) z^2 + 
  5 C^2 (-8 + 7 r) z^4 + C (30 - 29 r - 12 z) z^5\notag\\
  & + &
  2 z^3 (5 (-1 + r)^3 (-3 + 2 r) + z^4)\big)+\frac{z^2  }{ A ~B ~C^2   r^3} \big(-45 C^4 z^2 - 3 z^6 + C^2 z \big[18 + r (-54 + (54 - 17 r) r) - 45 z^3\big]\notag\\
  & +&
    C^3 \big[-3 + (-3 + r) r (-3 + 2 r) + 60 z^3\big] + C (B r z^2 + 18 z^5)\big)\Big]\Bigg\}.  
    	\end{eqnarray} 
In the following, we introduce the shorthand notations used to represent the relevant quantities: 
\begin{eqnarray}
	K &=&\frac{B}{A\,r} \Big(m_{Q}^2 \, C + m_{Q'}^2 \, r \Big),\nonumber\\
	D &=&\frac{C}{B^2} \Big((m_{Q}^2 \, C + m_{Q'}^2 \, r)\,B + s \,r\,z\,C \Big),\nonumber\\
	A&=& z (-1 + r + z) \, z,\nonumber\\
	B&=& r^2 + r (-1 + z) + (-1 + z) z,\nonumber\\
	C&=&1 -r.
\end{eqnarray}


\end{document}